\documentclass[a4paper,fleqn,usenatbib]{mnras}

\usepackage[T1]{fontenc}
\usepackage{ae,aecompl}

\usepackage{graphicx}	
\usepackage{amsmath}	
\usepackage{amssymb}	

\usepackage{newtxtext,newtxmath}

\newcommand{\se}[1]{\S\ref{sec:#1}}

\newcommand{\Fig}[1]{Figure~\ref{fig:#1}}

\newcommand{\be}{\begin{equation}}
\newcommand{\ee}{\end{equation}}
\newcommand{\bea}{\begin{eqnarray}}
\newcommand{\eea}{\end{eqnarray}}

\newcommand{\msun}{{\rm M}_\odot}

\newcommand{\Lsun}{L_\odot}
\newcommand{\ifm}[1]{\relax\ifmmode#1\else$\mathsurround=0pt #1$\fi}
\newcommand{\kms}{\ifmmode\,{\rm km}\,{\rm s}^{-1}\else km$\,$s$^{-1}$\fi}

\newcommand{\Mpc}{\,{\rm Mpc}}

\newcommand{\pc}{\,{\rm pc}}

\newcommand{\GyrI}{\,{\rm Gyr}^{-1}}
\newcommand{\sSFR}{\,{\rm sSFR}}
\newcommand{\am}{\mbox{\normalfont\AA}}

\newcommand{\ltsima}{$\; \buildrel < \over \sim \;$}
\newcommand{\lsim}{\lower.5ex\hbox{\ltsima}}
\newcommand{\gtsima}{$\; \buildrel > \over \sim \;$}
\newcommand{\gsim}{\lower.5ex\hbox{\gtsima}}

\def\la{\langle}

\def\Ms{M_*}
\def\Halpha{H$\alpha$}

\usepackage{color}

\title[FL IV]{FirstLight IV: Diversity in sub-L$_*$ galaxies at cosmic dawn }

\author[Ceverino et al.]
{Daniel Ceverino,$^{1,2,3}$\thanks{E-mail: daniel.ceverino@uam.es}
Michaela Hirschmann, $^{4}$
Ralf S. Klessen,$^{3,5}$ Simon C. O. Glover,$^{3}$
\newauthor
 St\'{e}phane Charlot$^{6}$,  Anna Feltre$^{7}$ 
\\ \\
$^{1}$Departamento de Fisica Teorica, Modulo 8, Facultad de Ciencias, Universidad Autonoma de Madrid, 28049 Madrid, Spain\\
$^{2}$CIAFF, Facultad de Ciencias, Universidad Autonoma de Madrid, 28049 Madrid, Spain \\
$^{3}$Universit\"{a}t Heidelberg, Zentrum f\"{u}r Astronomie, Institut
f\"{u}r Theoretische Astrophysik, Albert-Ueberle-Str. 2, 69120
Heidelberg, Germany\\
$^{4}$DARK, Niels Bohr Institute, University of Copenhagen, Lyngbyvej 2, 2100, Copenhagen $\mbox{\normalfont\O}$, Denmark \\
$^{5}$Universit\"{a}t Heidelberg, Interdisziplin\"{a}res Zentrum f\"{u}r Wissenschaftliches Rechnen, INF 205, 69120, Heidelberg, Germany \\
$^{6}$Sorbonne Universit\'e, CNRS, UMR7095, Institut d'Astrophysique de Paris, F-75014, Paris, France\\
$^{7}$INAF - Osservatorio di Astrofisica e Scienza dello Spazio di Bologna, Via P. Gobetti 93/3, 40129 Bologna, Italy}

\date{Accepted XXX. Received YYY; in original form ZZZ}

\pubyear{2021}

\begin{document}
\label{firstpage}
\pagerange{\pageref{firstpage}--\pageref{lastpage}}
\maketitle

\begin{abstract}
Using a large sample of sub-L$_*$ galaxies, with similar UV magnitudes, M$_{\rm UV}\simeq -19$ at $z\simeq6$, extracted from the FirstLight simulations, we show the diversity of galaxies at the end of the reionization epoch. We find
a factor $\sim$40 variation in the specific star-formation rate (sSFR).
This drives a $\sim$1 dex range in equivalent width of the [OIII]$\lambda$5007 line. 
Variations in nebular metallicity and ionization parameter within HII regions lead to a scatter in the equivalent widths and [OIII]/H$\alpha$ \ line ratio at a fixed sSFR.
 [OIII]-bright ([OIII]/\Halpha$>1$) emitters have higher ionization parameters and/or higher metallicities than \Halpha-bright ([OIII]/\Halpha$<1$) galaxies.
According to the surface brightness maps in both [OIII] and \Halpha,  [OIII]-bright emitters are more compact than \Halpha-bright galaxies.
\Halpha \ luminosity is higher than [OIII] if star formation is distributed over extended regions. 
[OIII] dominates if it is concentrated in compact clumps.
In both cases, the \Halpha-emitting gas is significantly more extended than [OIII]. 
\end{abstract}



\section{Introduction}

The current census of galaxies at the end of the reionization epoch, $z\simeq6$, has yielded a large number of sub-L$_*$ galaxies with relatively faint rest-frame UV magnitudes, M$_{UV}>-21$,  \citep{Stark16}.
They represent the majority of galaxies at these high redshifts and 
they provide the bulk of the photons responsible for the reionization of the Universe \citep{Robertson13, Naidu20}. 
However, little is known about their basic properties: mass, star formation (SF), size, or physical conditions in the interstellar medium (ISM).

Observations of these sub-L$_*$ galaxies show steep UV slopes \citep{Bouwens03, Bouwens06, Stanway05}, indicating young and metal poor stars. The modelling of their photometry  hints at low masses, $\Ms \leq 10^9 \ \msun$, and high sSFR values around $3-10 \GyrI$ \citep{Stark09, Salmon15}.
However, are these typical values? How diverse is the population of sub-L$_*$ galaxies at the end of reionization?
Understanding  the origin of this diversity will allow us to design efficient observations that look for galaxies with different properties.  The {\it James Webb Space Telescope} ({\it JWST}) and the next generation ground-based facilities will soon explore this mostly uncharted territory.

Available observations of high-redshift galaxies show hints of bright emission in rest-frame visible lines, such as [OIII]$\lambda$5007, H$\beta$ or \Halpha \ \citep{Chary05, Stark13, Smit16, Rasappu16, Faisst16, DeBarros19}.
Some galaxies exhibit extreme nebular conditions, with very high  equivalent widths,  EW([OIII]$\lambda\lambda4959, 5007 + H\beta) \geq 1000 \ \am$ \citep{Smit14, Smit15, RobertsBorsani16, Endsley21}, although median values are around 600 $\am$ \citep{Labbe13}.
However, it is not clear what drives these high values or whether these galaxies with high EW$\geq 100 \ \am$  are representative examples of the underlying population of sub-L$_*$ galaxies.

{\it JWST} will give us the opportunity to unveil the diversity of galaxies at reionization.
In particular, NIRSpec spectroscopy will disentangle the individual lines in sub-L$_*$ galaxies at $z\geq6$, particularly in the wavelength ranges around [OIII]$\lambda5007$ and \Halpha, two of the brightest lines in galaxy spectra at high z \citep{Stark16}.
Its IFU capability will also constrain the spatial extent of these emission regions.
These observations will tell us whether galaxies with similar rest-frame UV properties have also similar rest-frame visible lines. They will unveil the diversity in line strengths.
This can give us clues about the main drivers of the observed variations.
Do galaxies with different  [OIII]/\Halpha \ ratios differ in 
 any global property or is it only related to the local conditions within HII regions?
Are the properties of [OIII]-bright ([OIII]/\Halpha$>1$) galaxies different from the properties of [OIII]-faint emitters?
{\it JWST} may tell us whether there are different populations of line emitters at high z. 

Cosmological simulations of galaxy formation can give us a first insight into this galaxy diversity at cosmic dawn \citep{OShea15, PaperI, Katz19, Ma18, Pallottini19}.
However, this requires large and unbiased samples of simulated galaxies.
The FirstLight database is a mass-limited sample of simulated galaxies at cosmic dawn \citep{PaperI}.
It successfully encompasses a large diversity of complex SF histories, characterised by frequent SF bursts of different strength and duration \citep{PaperII}. 
This translates into a large diversity of spectral energy distributions \citep{PaperIII}, consistent with current observations. The rest-frame UV magnitudes range from M$_{\rm UV}=-12$ to $-22$ and the stellar masses range from $\Ms=10^6$ to $10^{9.5} \ \msun$, 
with a relatively large number of galaxies with similar UV magnitudes but different stellar masses.

This paper uses the FirstLight database and it focuses on a narrow range of UV magnitudes: M$_{\rm UV}\simeq -19$ at $z\simeq6$.
This allows us to study the galaxy diversity almost independently of the trends from galaxy scaling relations.
We aim to understand the physical origin of variations in the strength of 
two of the brightest rest-frame visible lines,
 [OIII]$\lambda5007$ (hereafter [OIII]) and \Halpha.
 According to the line fluxes of FirstLight galaxies described in \cite{MIRI},  
 the median flux is about a few times $10^{-18} \ {\rm erg} \ {\rm s}^{-1} {\rm cm}^{-2}$.  
 The whole sample with M$_{\rm UV}\simeq -19$ at $z\simeq6$ could be detectable by NIRSpec with exposure times that range between 2 and 0.5 hours  per target. 
 Another comparison between FirstLight and observations indicates that
their steep UV slopes are consistent with little or no dust attenuation \citep{Bouwens16b, PaperIII}. 
Therefore, the selected UV magnitude bin is ideal to study the diversity in line strengths without the complications implied by dust attenuation and other selection effects.

The outline of this paper is as follows.
Section \se{runs} presents the  sample selection from the FirstLight simulations and the post-processing models used to generate the nebular emission lines.
The results section (\se{R}) includes some examples of galaxies with different [OIII] equivalent widths  (\se{examples}), the diversity of [OIII]/\Halpha \ line ratios  (\se{lines}), and the spatial extent of this emission  (\se{size}).
Finally, section \se{summary} ends the paper with the summary and discussion.

\section{Methods}
\label{sec:runs}

\subsection{Sample selection}

The parent sample is a complete mass-selected subsample from the FirstLight database of simulated galaxies, described fully in \cite{PaperI}.
The subsample contains 290 halos with a maximum circular velocity ($V_\mathrm{max}$) between 50 and 200 $\kms$, selected at $z=5$.
The halo mass range is between a few times $10^9$ and $10^{11} \ \msun$.
It does not include more massive and rare halos with number densities lower than $\sim 3 \times 10^{-3} (h^{-1} \Mpc)^{-3}$.
It also excludes small halos in which galaxy formation is not sufficiently efficient.

From the parent sample, a complete luminosity-selected sample was selected based on  two criteria: a redshift around $z\simeq6$ ($z=5.5-6.5$) and an absolute UV magnitude at $1500 \ \am$ of $M_{UV} \simeq -19$ (between $-19$ and $-19.5$ magnitudes). 
These two criteria give a sample of 150 snapshots of sub-L$_*$ galaxies with observed H band magnitudes between 27 and 28 at the end of the reionization epoch. On average, there are four snapshots per distinct galaxy with a total number of 35 galaxies. Due to the burstiness of the SF histories \citep{PaperII}, all snapshots can be considered as different galaxies without introducing any significant bias in the results. 

\subsection{The FirstLight Simulations}
  
 The simulations are performed with the  $N$-body+Hydro \textsc{ART} code
\citep{Kravtsov97,Kravtsov03, Ceverino09, Ceverino14, PaperI}.
Gravity and hydrodynamics are solved by an Eulerian, adaptive mesh refinement (AMR) approach.
The code includes  astrophysical processes relevant for galaxy formation, such as gas cooling by hydrogen, helium and metals.
Photoionization heating uses a uniform cosmological UV background with partial self-shielding. 

Star formation and feedback (thermal+kinetic+radiative) models are described in \cite{PaperI}.
The SF timestep is set to 5 Myr, although the dynamics is solved using an adaptive time stepping with a maximum temporal resolution of $\sim$1 Kyr.
The simulations follow metals from  supernovae type-II and type Ia, using yields that approximate the  results from \cite{WoosleyWeaver95}, as described in  \cite{Kravtsov03}.  
We assume that the effect of AGN feedback on these  sub-L$_*$ galaxies is very minor and it is not included.
 The DM particle mass resolution is $m_{\mathrm{DM}} = 10^4 \ \msun$. The minimum mass of star particles is $100 \ \msun$.
 The maximum spatial resolution is always between 8.7 and 17 proper pc (a comoving resolution of 109 pc after $z<11$).
 \cite{PaperI} includes more details about the FirstLight simulations.

\subsection{Post-processing}

Global galaxy properties, such as stellar mass, UV magnitude or slope ($\beta$) are extracted from the FirstLight database. 
In summary, the continuum emission of the simulated galaxies is computed using publicly available tables from the Binary Population and Spectral Synthesis (BPASS) model \citep{Eldridge17} including nebular continuum. More details can be found in \cite{PaperIII}. 

The emission-line maps are computed in two steps.
First, a uniform 6x6 kpc$^{2}$ grid is laid at the face-on view of each galaxy.
The 2D grid is centered with respect to the stellar distribution.
The face-on view is defined using the angular momentum of the cold gas.
Each 100-pc-wide pixel stores the mass in stars younger than 10 Myr, the average gas density and metallicity of the warm gas ($T<5 \times 10^4 {\rm K}$).
The use of different pixel sizes from 50 to 300 pc gives similar overall results because the typical cell size of the simulation ($\sim10 \ \pc$) is much smaller than the pixel size.

\begin{figure*}
	\includegraphics[width=0.649 \columnwidth]{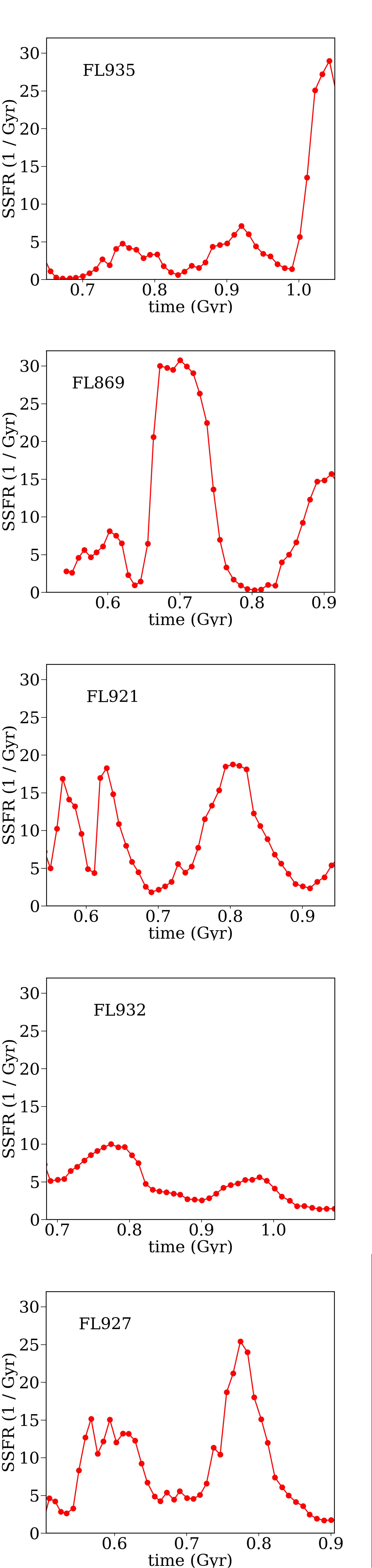}
	\includegraphics[width=1.1\columnwidth]{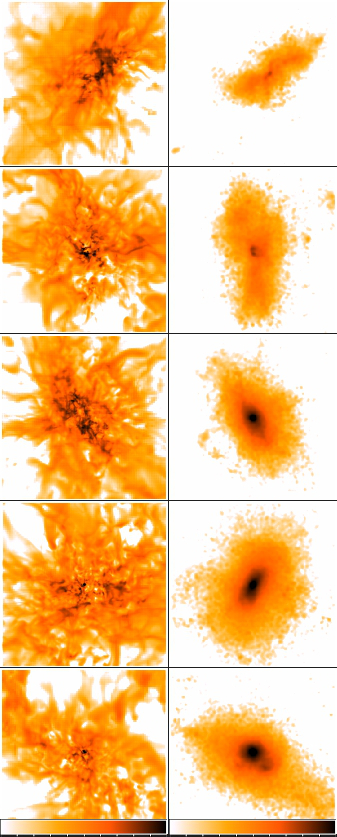}

		 \caption{Five examples of sub-L$_*$ galaxies at $z\simeq6$. Each row shows projections of stars (right), gas (middle) surface density and the sSFR evolution (left) of the main galaxy progenitor during the previous $\sim$400 Myr. The size is 10x10 kpc$^2$. The color bars range from 1 to 500 $\msun {\rm pc}^{-2}$  in 10 log-scaled ticks. }
	  \label{fig:examples}
\end{figure*}
\begin{figure*}
	\includegraphics[width= \columnwidth]{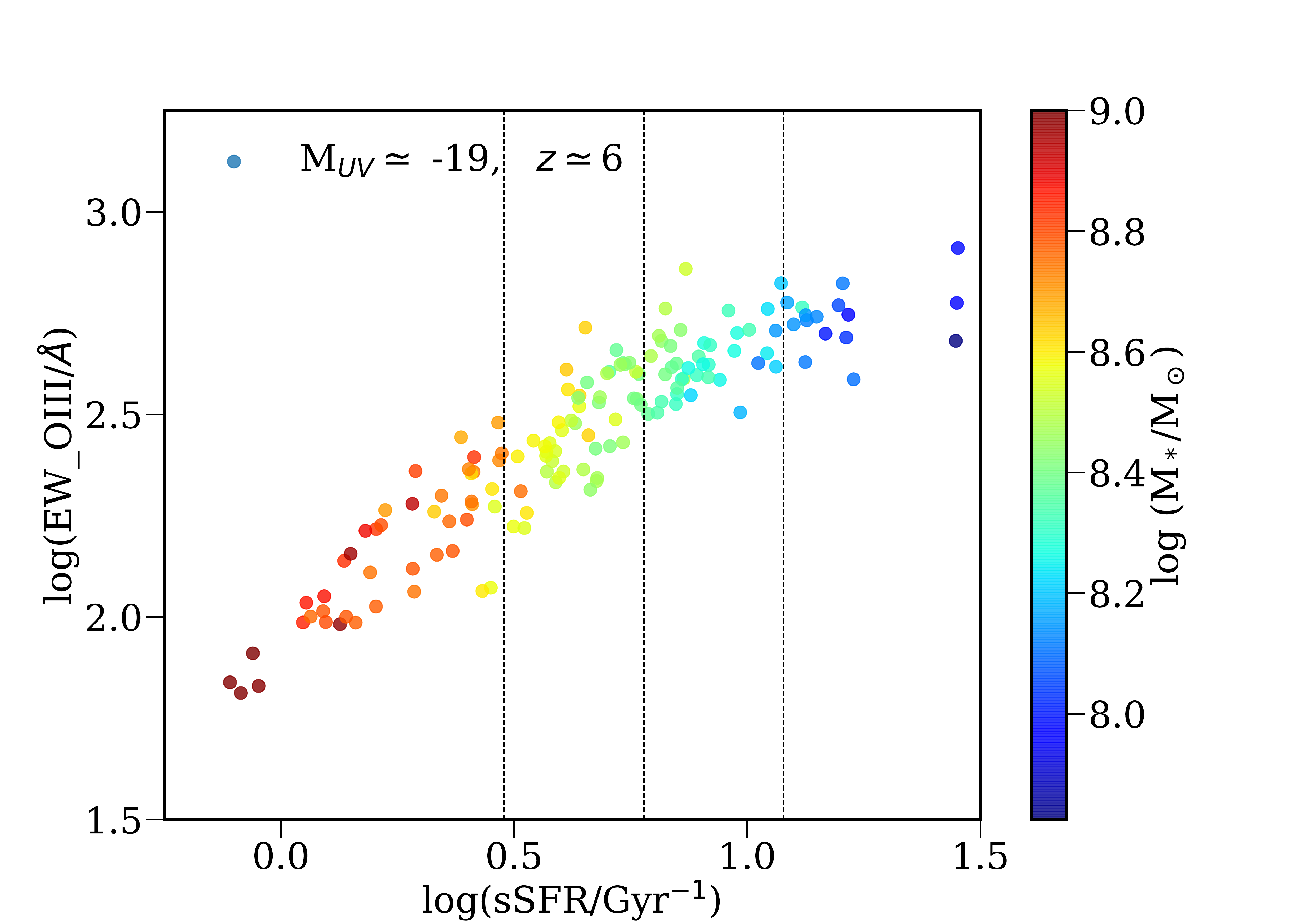}
	\includegraphics[width= \columnwidth]{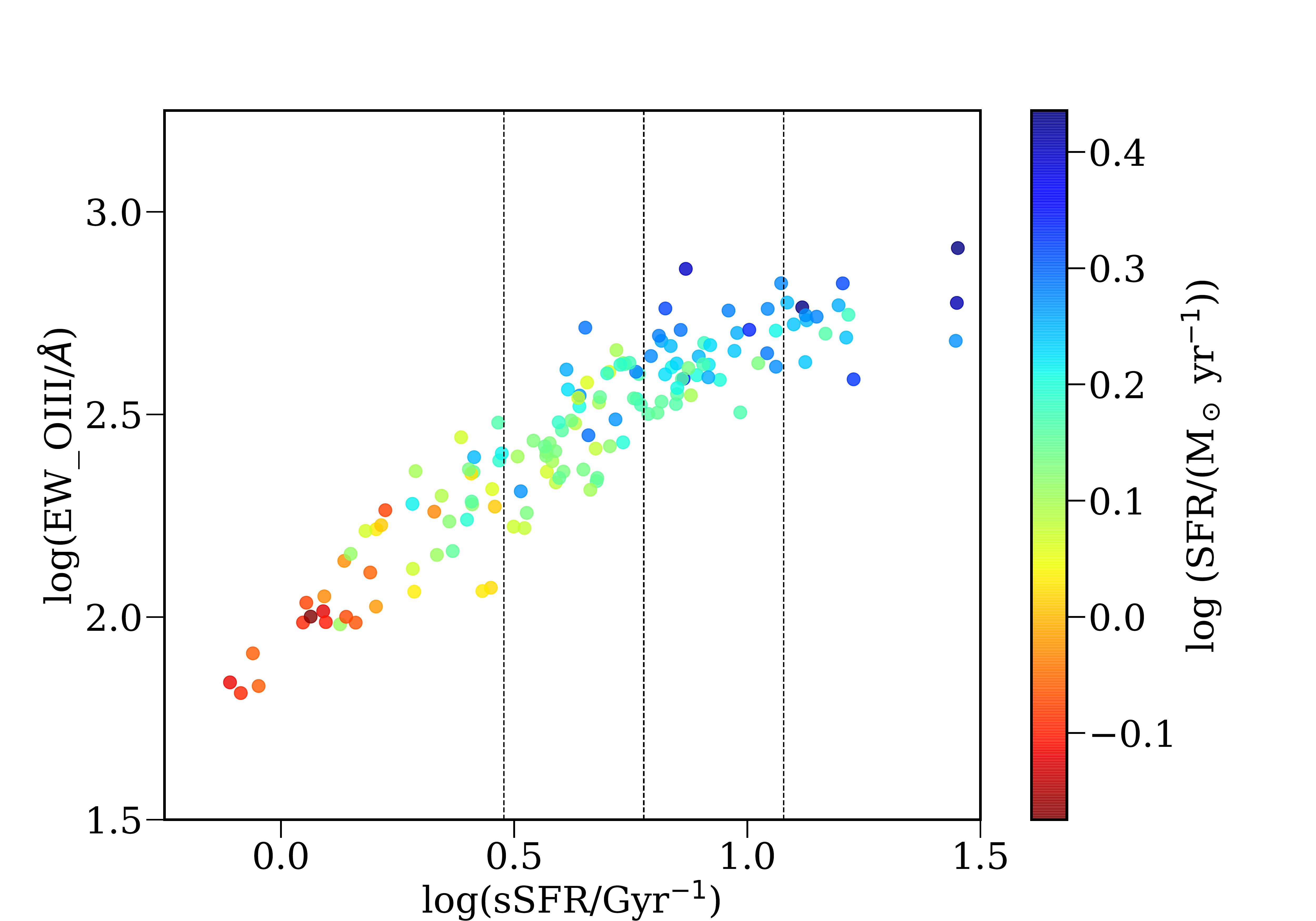}	
	\includegraphics[width= \columnwidth]{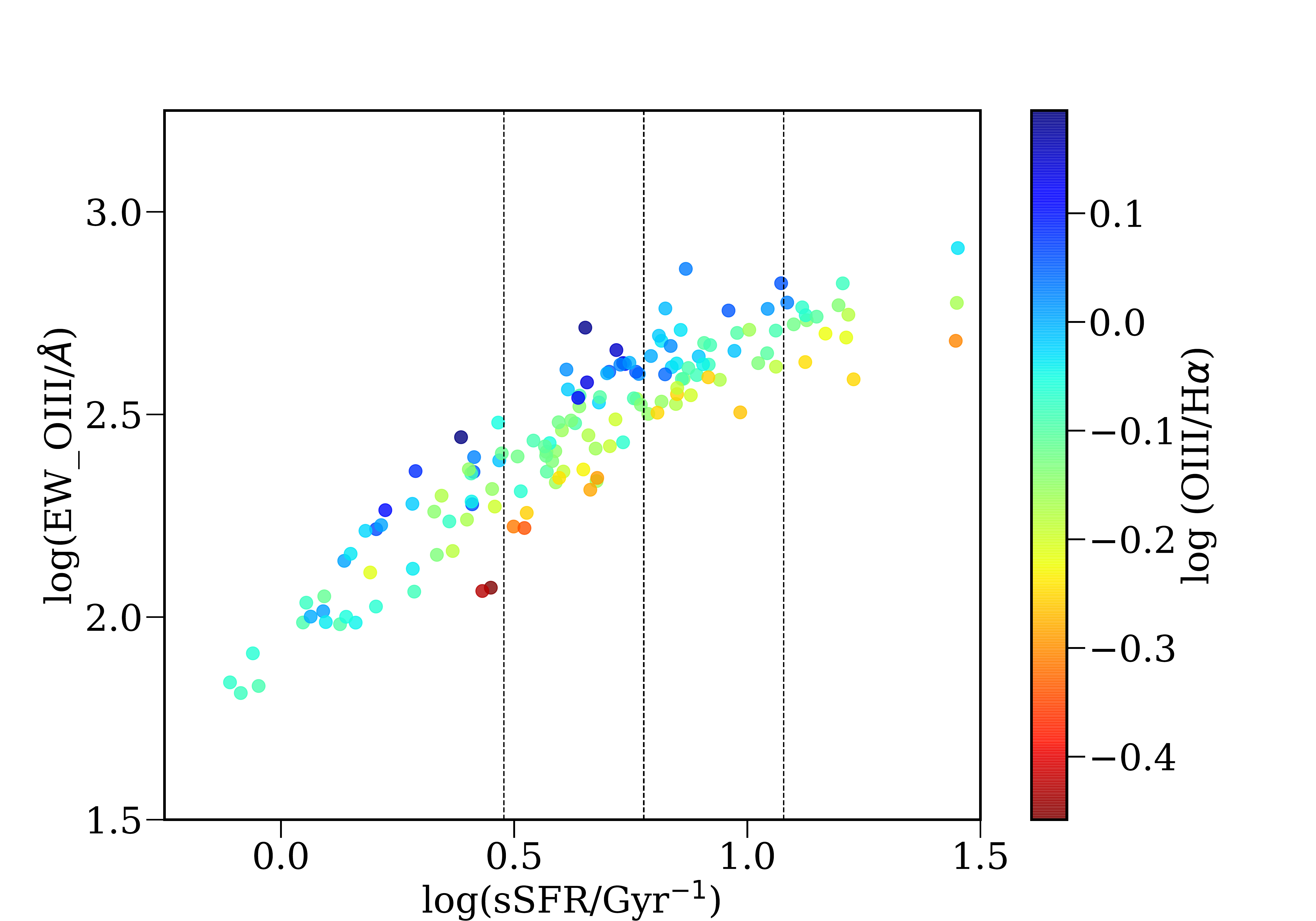}	
	\includegraphics[width=\columnwidth]{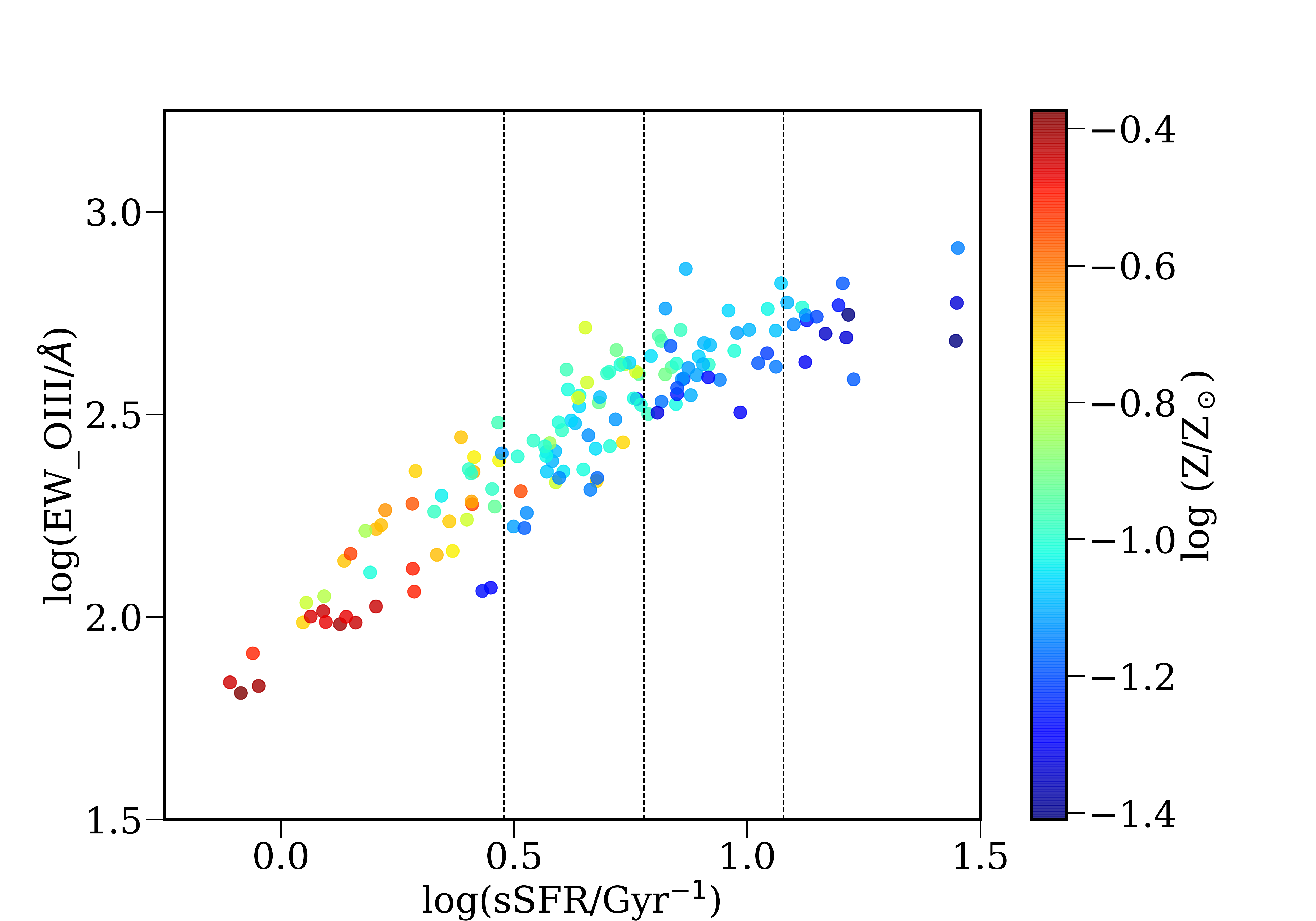}
	\includegraphics[width=\columnwidth]{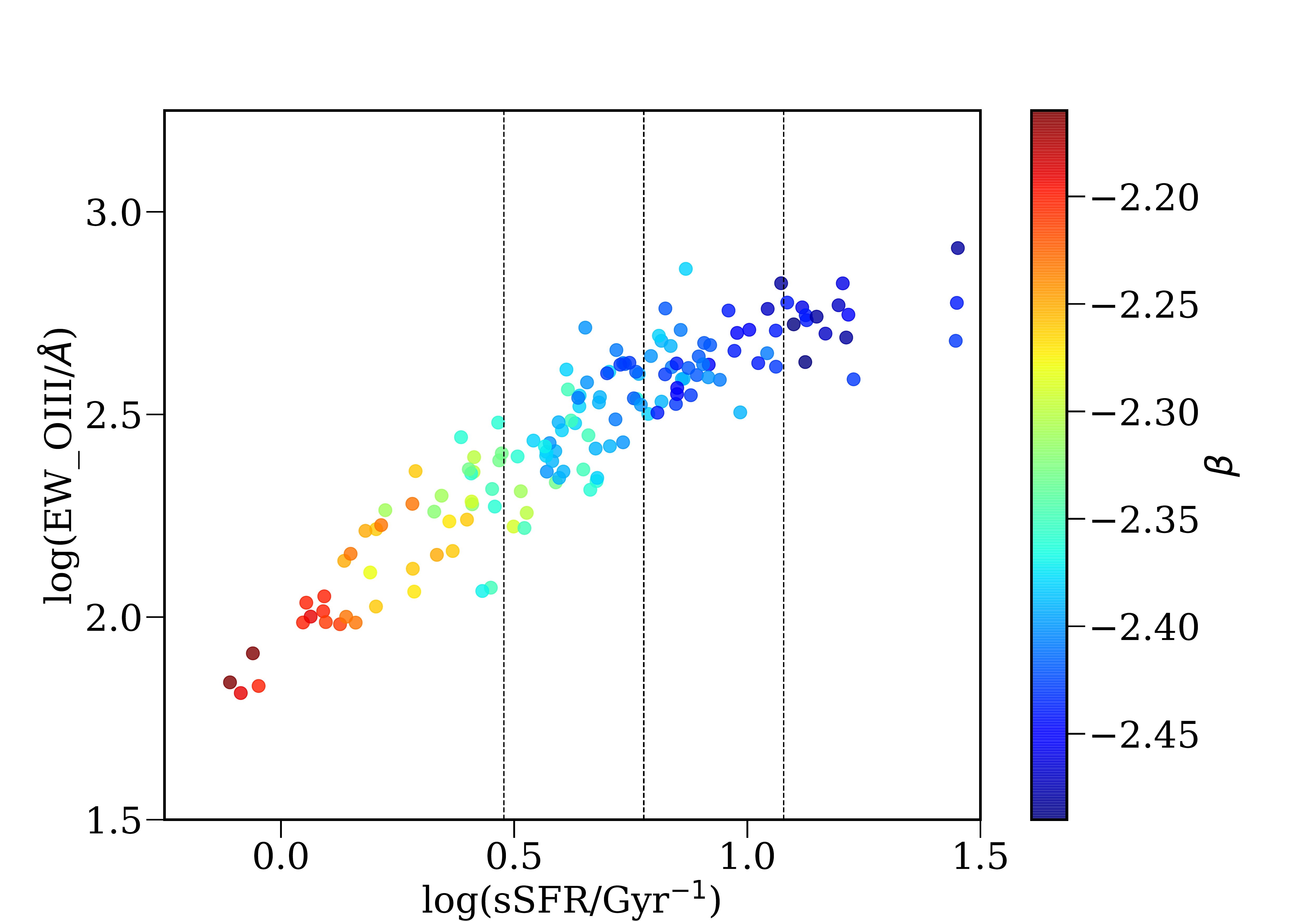}	
	\includegraphics[width=\columnwidth]{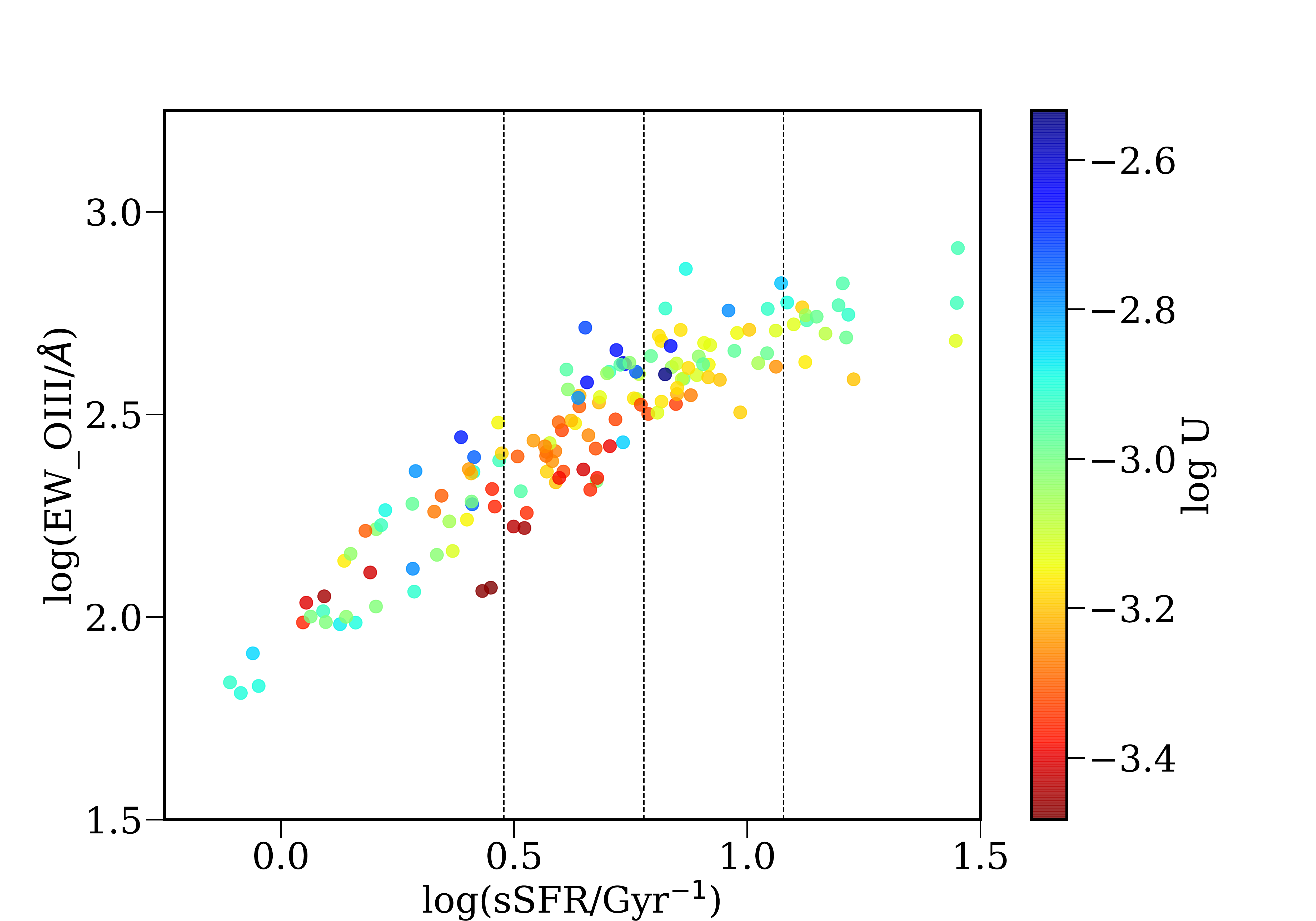}	
		 \caption{Equivalent width (EW) of the  [OIII]$\lambda5007$ line versus specific star formation rate. The points are coloured by stellar mass, star formation rate, [OIII]/H$\alpha$ luminosity ratio, UV slope, SFR-weighted nebular metallicity, and SFR-weighted ionization parameter. The vertical lines mark the position of the star-forming main sequence and its scatter \citep{PaperII}. The EW increases with stellar mass and SFR  but the scatter at fixed sSFR is correlated with the line luminosity  ratio and it is driven by the ionization parameter.
		 }
	  \label{fig:EW}
\end{figure*}

The gas and star properties of the individual pixels are then used to compute emission lines based on the methodology and emission line models for young stellar populations described in \cite{Hirschmann17, Hirschmann19}. Specifically, we adopt the grid of nebular-emission models of star-forming galaxies computed by \cite{Gutkin16}. These calculations combine the latest version of the \cite{BC03} stellar population synthesis model with Cloudy \citep{Ferland13}, following the method outlined by \cite{Charlot01} for each element of a simulated galaxy. This implies that each pixel is composed of nebular emission emerging from
a steady population of ionization-bounded HII regions.
This is a reasonable assumption for modeling line emission, 
since the majority of ionizing photons are released during the first 10 Myr of evolution of a single stellar population \citep{Gutkin16}.
This is just two time-steps of SF in the simulation, in which the SFR is roughly constant in most cases.

The emission line grid, table 1 in \cite{Hirschmann17} and table 3 in \cite{Gutkin16}, includes models in wide ranges of interstellar (gas+dust) metallicities, ionization parameters, dust-to-metal mass ratios, HII-region densities and carbon-to-oxygen abundance ratios.
With each galaxy pixel, we associate the SF emission-line model from the Gutkin grid (described above) with closest
pixel-average values. 
We select the grid metallicity closest to the average metallicity in a given pixel.
The ionization parameter, log U, is computed based on equation 1 in  \cite{Hirschmann17} using the instantaneous  SFR and the filling factor = $n_{\rm gas}/n_{\rm H}$, where $n_{\rm gas}$ is the density in warm gas and $n_{\rm H}$ is the hydrogen density within a HII region.
 For the dust-to-metal mass ratio, the C/O ratio and the hydrogen density, we assume fixed values (0.3, solar C/O and $n_{\rm H}=100 \ {\rm cm}^{-3}$), as these quantities are either not modelled or not resolved.
We assume that the emission from old stellar populations and non-stellar sources is not significant in these actively star-forming and young galaxies.

The galaxy [OIII] and \Halpha \ luminosities given by this model are very similar to the ones discussed by \cite{PaperIII}, who used the BPASS model \citep{Eldridge17,XiaoStanway18}.
BPASS yielded a factor 0.8 lower luminosity than the ones used in this paper.
The [OIII]/\Halpha \ ratios agree to within 10\%.  This assures us that the results reported in this paper are independent of a particular emission-line model.

\section{Results}
\label{sec:R}

\subsection{Extreme versus typical galaxies}
\label{sec:examples}

The sample of sub-L$_*$ galaxies with similar UV luminosities and redshifts ($M_{UV} \simeq -19$ at $z\simeq6$) show a large diversity. 
The median values are $\Ms = 10^{8.5} \ \msun$, ${\rm SFR}=1.4 \ \msun$ yr$^{-1}$ and $\sSFR = 5 \GyrI $,
but the sample ranges over $\Ms = 10^{7.8} - 10^9 \ \msun$ and ${\rm SFR}=0.67- 2.7 \ \msun$ yr$^{-1}$.
These values are consistent with observations \citep{Stark13, Duncan14, Salmon15, Song16} as described in \cite{PaperII, PaperIII}.
\Fig{examples} shows five examples of this diversity.
The top row shows a galaxy at the peak of an extreme burst of star formation, as shown by its sSFR history. 
Its value, $\sSFR = 30 \GyrI$, is a factor 6 higher than the median value of the whole sample.
The burst is driven by a multiple merger.
This boosts  its UV luminosity in spite of its lower-than-average stellar mass, $\Ms=10^8 \ \msun$. 
The gas is distributed in a few dense and compact clumps that concentrate most of the star formation.
The second row displays another example of a merger-driven burst, but the peak is more representative of a typical burst, $\sSFR=16 \GyrI$ \citep{PaperII}.
This example shows a compact gas distribution, due to the final merger coalescence. 

The third row shows a typical, main-sequence star-forming galaxy at $z\simeq6$ with $\sSFR=5 \GyrI$.
According to its sSFR history, the galaxy is not in a maximum or minimum value, but somewhere in between.
The gas is significantly more extended than in the previous cases and it shows smaller clumps distributed throughout the galaxy.
The stellar mass,  $\Ms=4 \times10^8 \ \msun$,  is higher than in the previous starburst examples and the stellar distribution shows a dense center. 
This is  an indication of a mature population, formed in multiple bursts in the past 200-300 Myr.
The fourth row illustrates  a galaxy that lies a factor of 2 below the SF main-sequence.
 Its gas distribution looks more diffuse and less clumpy than in the other cases.
The fifth row provides an example of a quiescent, post-starburst galaxy (sSFR$ \ =1.7 \GyrI$), 100 Myr after the last burst of star formation.
Its stellar mass, $\Ms=6 \times 10^8 \ \msun$,  is significantly higher than average and it is mostly concentrated in  a dense stellar center, with non-clumpy gas around it.
In conclusion, there is a factor 40 variation in sSFRs even within galaxies with similar UV luminosities.
This drives a large variety in stellar and gas distributions.

This galaxy diversity can also be seen in the strength of emission lines.
\Fig{EW} shows the equivalent width (EW) of the [OIII] line for all galaxies of the sample.
The EW increases with sSFR, reaching EW$\simeq$1000 $\am$ for extreme SF bursts with the highest sSFR, in agreement with observations \citep{ Smit14, Smit15, RobertsBorsani16, Endsley21}.
However, a typical example of a galaxy on the star-forming sequence (FL921 in the third row of \Fig{examples}) shows a moderate EW$\simeq$ 300 $\am$.
The most massive galaxies with the lowest SFR have the lowest EW $\leq$ 100 $\am$ (FL927 at the bottom row of \Fig{examples}). 
This is a factor $\sim$10 lower than the extreme SF bursts.
This range is partially due to the 1 dex variation in stellar mass of galaxies with similar UV luminosity \citep{PaperIII} and the factor 4 difference in SFR.
Both the UV  slope and the SFR-weighted, nebular metallicity ($Z$) exhibit a similar behaviour, driven by the diversity in stellar mass \citep{Langan20} and SFR.

There is a significant, intrinsic dispersion in the above relation between EW and sSFR 
of $\sigma \simeq 0.1 $ dex.
Therefore,  there are other drivers of high EWs.
At a fixed sSFR, galaxies with higher EW also have higher [OIII]/H$\alpha$ luminosity ratio and higher SFR-weighted, ionization parameters (\Fig{EW}).
This indicates that the nebular conditions,
such as gas and SFR densities,
drive this scatter.
Even between galaxies with similar UV magnitudes, sSFR, and redshifts we expect different ISM conditions.

\subsection{Line ratios}
\label{sec:lines}

\begin{figure}
	\includegraphics[width=\columnwidth]{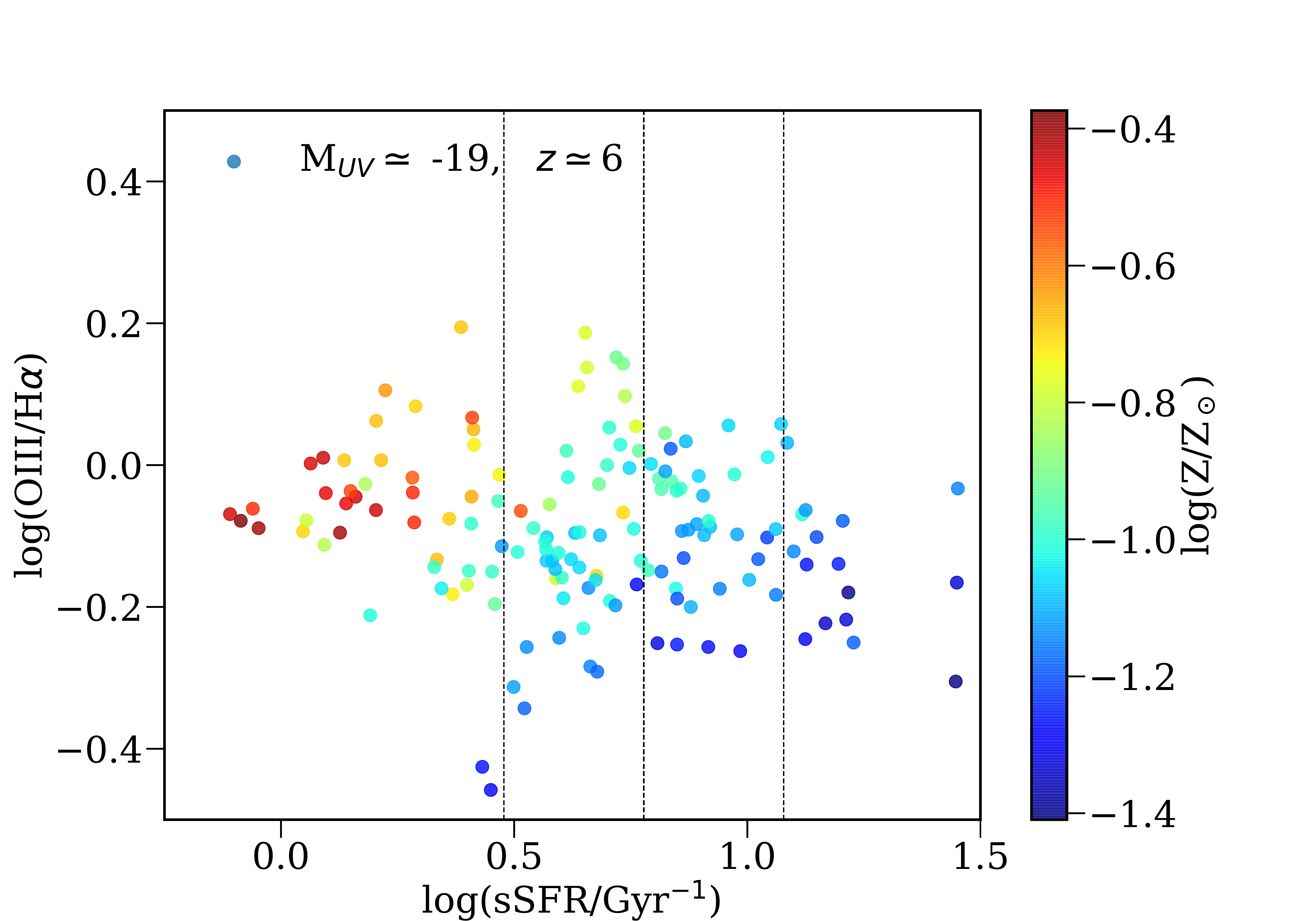}
	\includegraphics[width=\columnwidth]{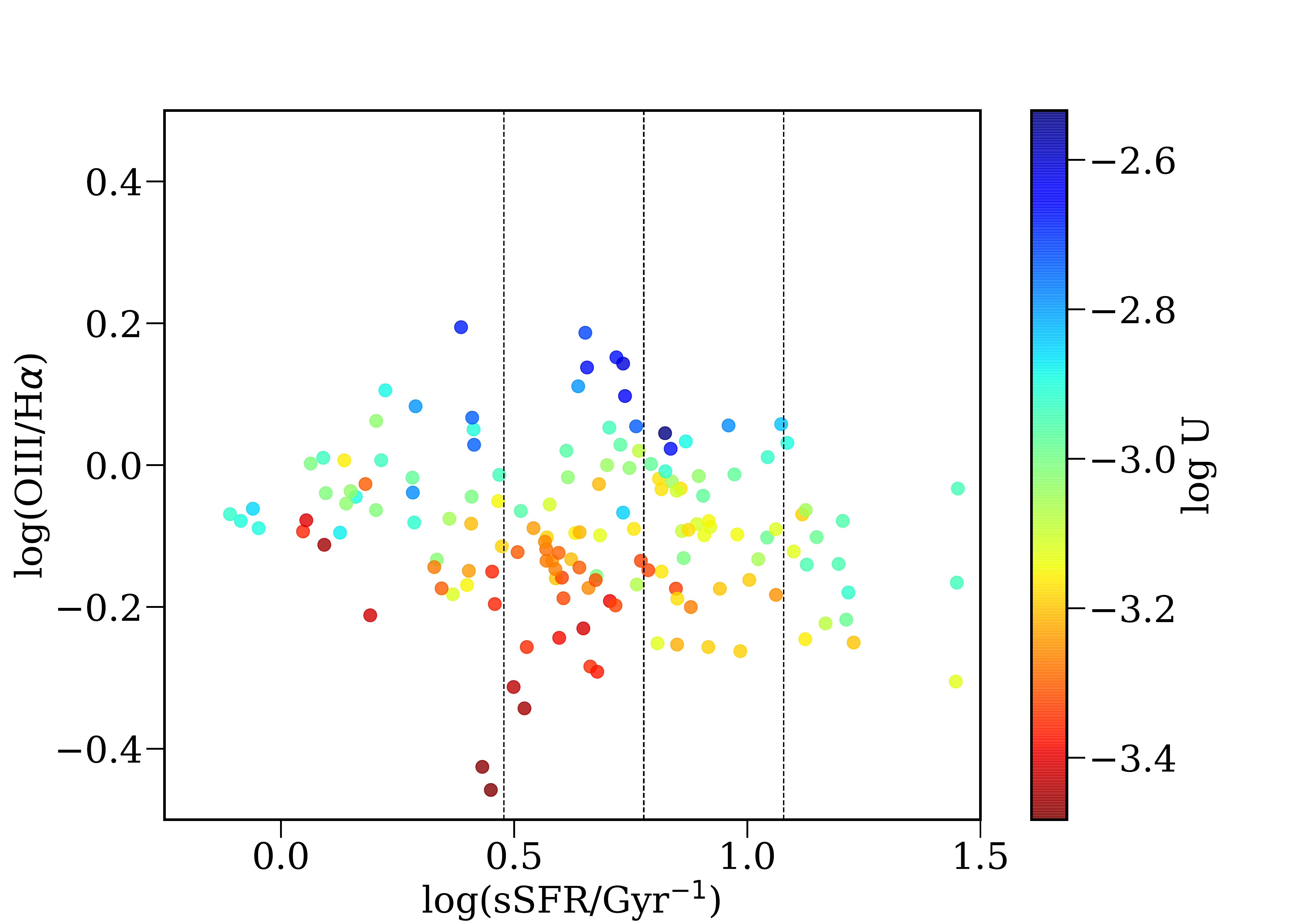}	
		 \caption{[OIII]/H$\alpha$ line luminosity ratio versus sSFR. 
		 The points are coloured by SFR-weighted  nebular metallicity (top) and ionization parameter (bottom). 
		 The vertical lines mark the position of the star-forming main sequence and its scatter. 
		 At a fixed sSFR, high [OIII]/H$\alpha$  ratios are driven by a high metallicity and/or a high ionization parameter.}
	  \label{fig:OHalpha}
\end{figure}
\begin{figure*}
	\includegraphics[width=0.9 \columnwidth]{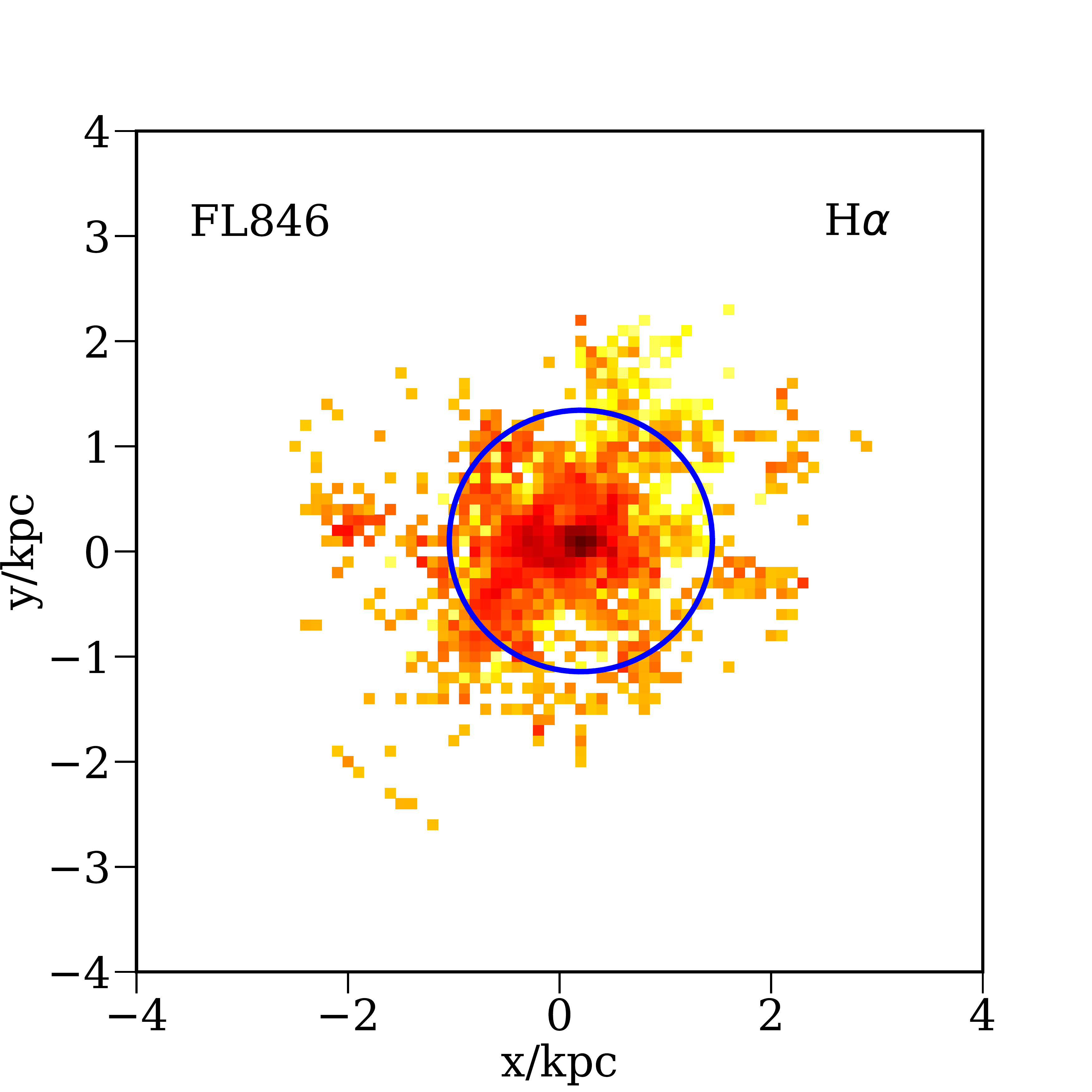}	
	\includegraphics[width=0.9 \columnwidth]{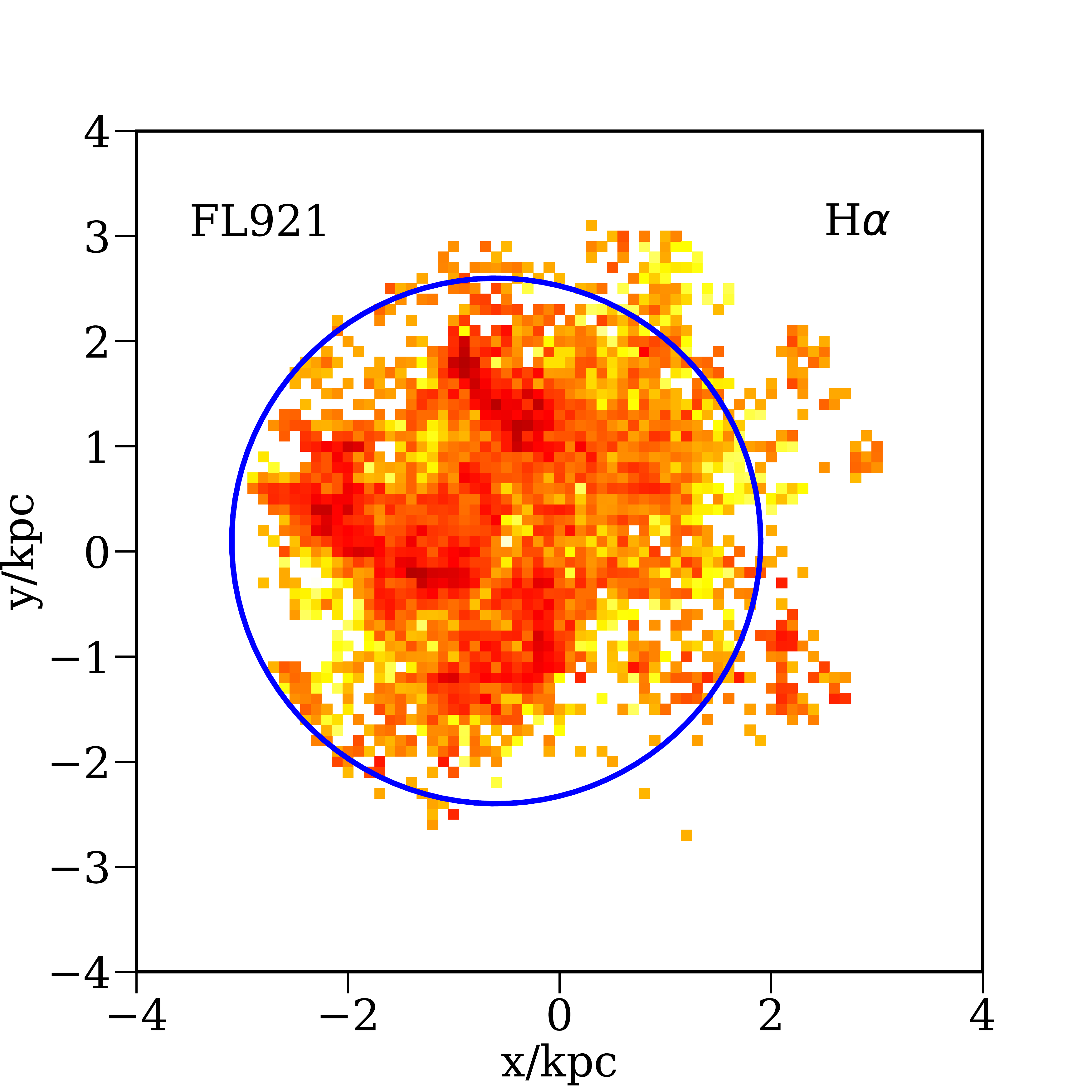}
	\includegraphics[width=0.9 \columnwidth]{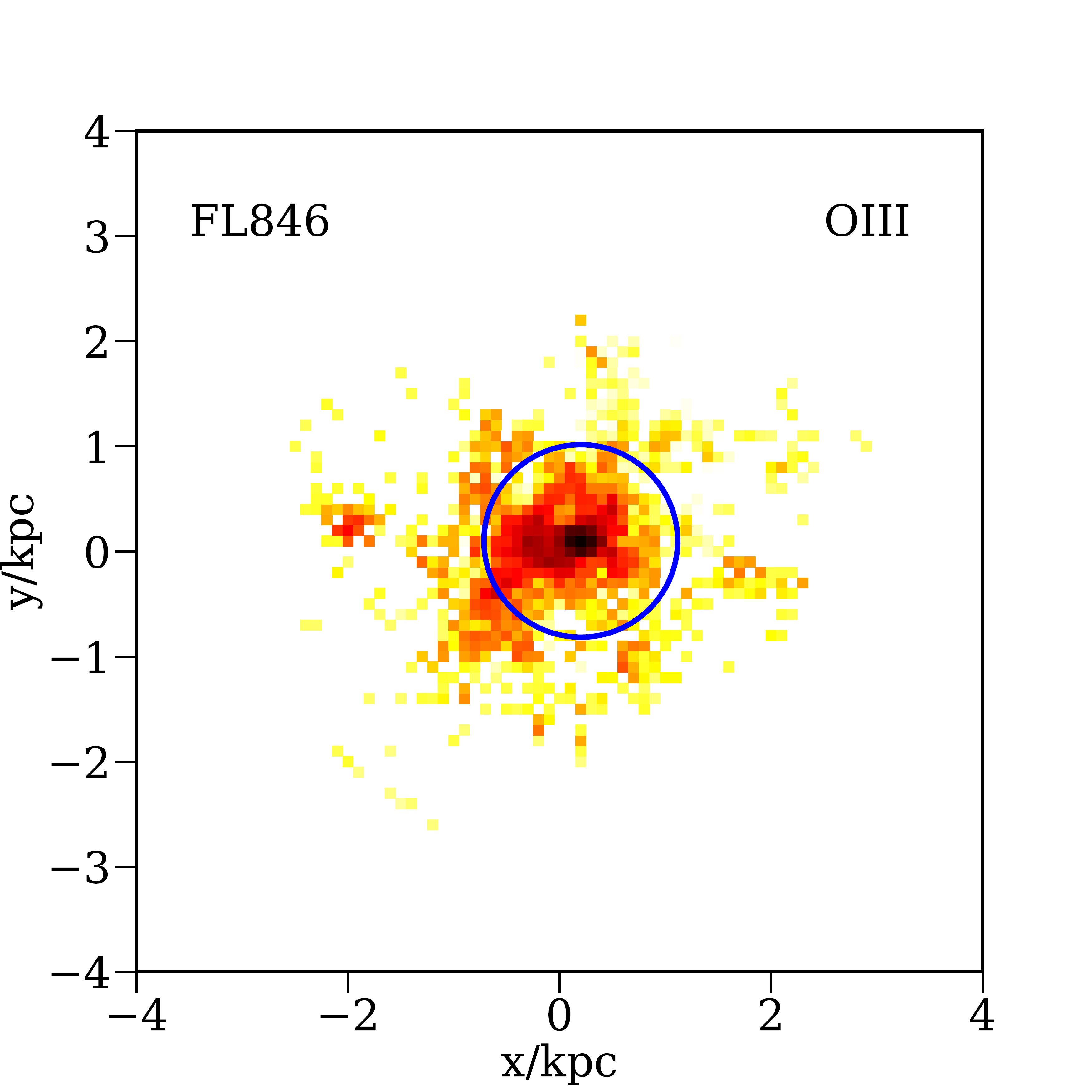}
	\includegraphics[width=0.9 \columnwidth]{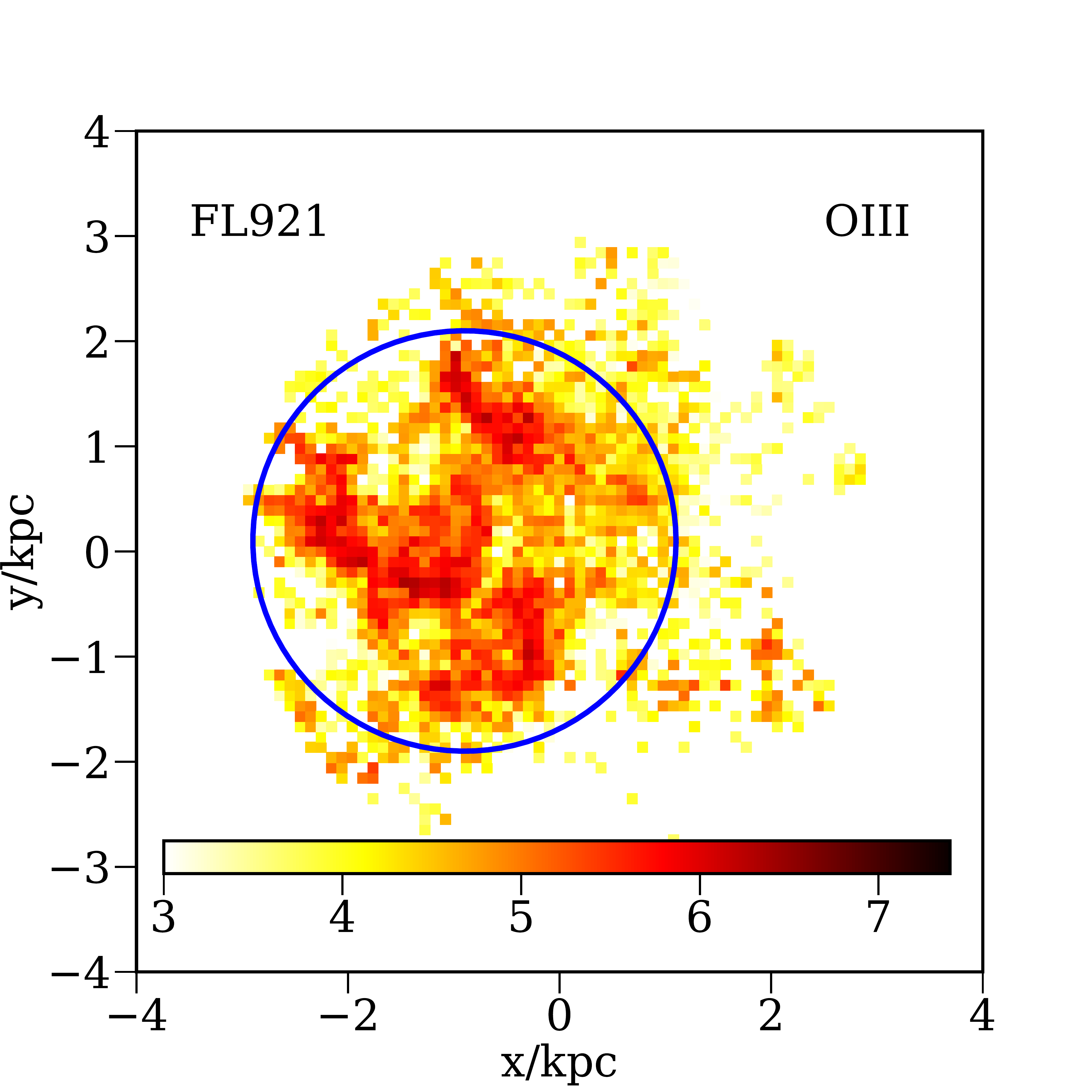}	
	\includegraphics[width=0.9 \columnwidth]{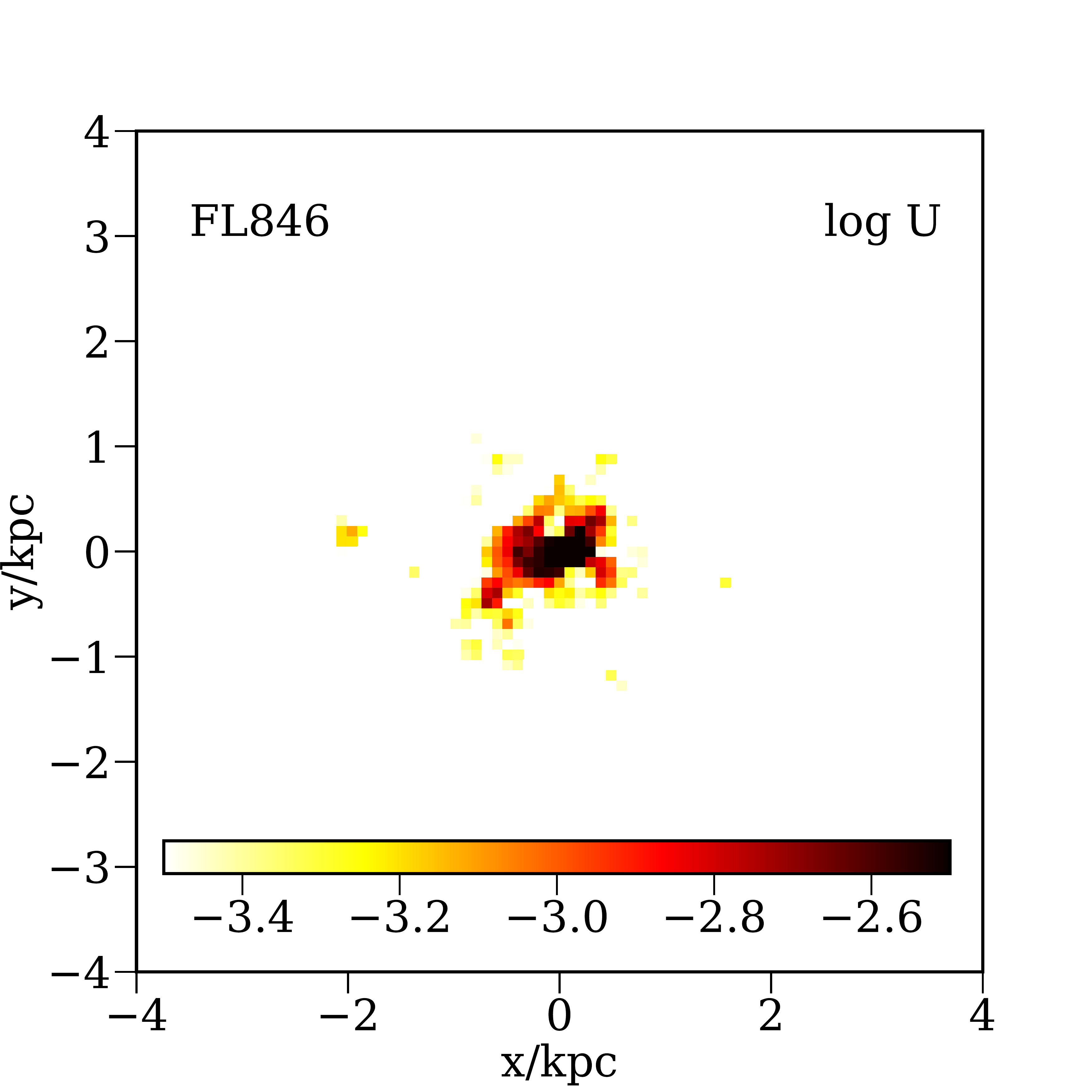}
	\includegraphics[width=0.9 \columnwidth]{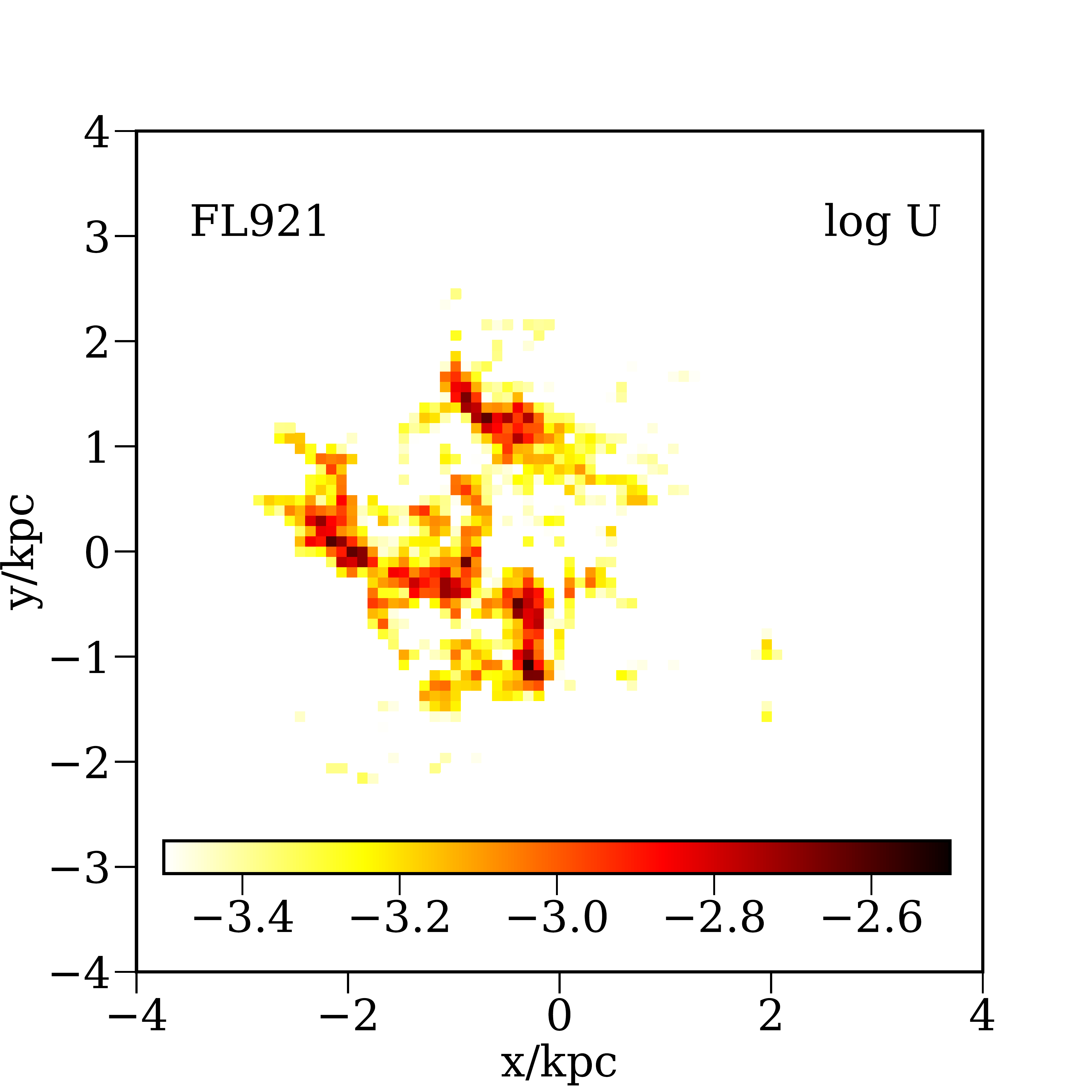}	
		 \caption{ Maps of \Halpha \ (top),  [OIII] luminosity (middle), and ionization parameter (bottom) 
		 for two galaxies with similar sSFR but different line ratios:  log([OIII]/\Halpha$)=0.15$ (left) and  log([OIII]/\Halpha$)=-0.17$ (right). 
		 Blue circles mark 2 times the half-light radius. 
		 The colorbar indicates the line luminosity in log($\Lsun$/pixel),
		 where the pixel size is 100 pc. 
		 Galaxies with high line ratios have systematically more compact SF regions.  
		 The distribution in \Halpha \ is significantly more extended than that in [OIII].}
	  \label{fig:maps}
\end{figure*}

The FirstLight simulations indicate a diversity in the conditions of star-forming regions at high z, 
such as variations in ionization parameter.
This translates into different [OIII]/\Halpha \ luminosity ratios even between galaxies with similar sSFR.
For example, some galaxies are [OIII]-bright,  log([OIII]/\Halpha$)>0$, with [OIII] luminosities up to a factor of 2 higher than \Halpha.
They are analogous to rare compact emission-line galaxies at $z\sim0$ \citep{Izotov11} or extreme [OIII]+H$\beta$ emitters at $z\sim2-3$ \citep{Forrest17, Tang21}. At higher redshifts, these kind of galaxies may be more frequent. 
Within the FirstLight sample around the main-sequence, 20\% of them have higher luminosities in [OIII]  than in \Halpha. 
Therefore, they constitute a significant fraction of the galaxy population at the end of reionization.

\Halpha-bright emitters, log([OIII]/\Halpha$)<0$, dominate the FirstLight sample. They have 
 \Halpha \ luminosities up to a factor of 3 higher than [OIII]. 
 However, they are currently getting little attention mostly because \Halpha \ becomes inaccessible for NIRSpec spectroscopy at $z>6.5$. 
Longer wavelengths (MIRI) will be needed to unveil this population at higher redshifts \citep{MIRI}
but they are an important galaxy population at the end of reionization, $z<6.5$.

This diversity in  [OIII]/\Halpha \ ratio depends primarily on the ionization parameter and secondarily on nebular metallicity (\Fig{OHalpha}). 
In our sample of low-metallicity galaxies, an increase in metallicity leads to higher line ratios, in a way similar to the [OIII]/H$\beta$ ratio \citep{Langan20}. 
This is due to the low values of metallicity, log(Z/Z$_\odot)<-$0.4, which place most of these galaxies in the ascending part of the [OIII]/H$\beta$-Z curve \citep{Gutkin16}. The appendix shows the [OIII]/H$\beta$ ratio for comparison.
However, metallicity also depends on stellar mass, and therefore, 
in this UV-selected sample, galaxies with higher-than-average mass and metallicity
have lower-than-average sSFR but also relatively high line ratios. 
The opposite is true for the metal-poor, less-massive starbursts.
Consequently, the trend between line ratio and sSFR is largely due to metallicity effects.
Around the main sequence, more metal-rich galaxies have higher line ratios than more metal-poor galaxies with the same sSFR.

The ionization parameter is  the main driver of the scatter of the [OIII]/\Halpha \ ratio at a fixed sSFR.
[OIII]-bright galaxies have higher ionization parameters than \Halpha-bright galaxies.
Interestingly, these [OIII] emitters do not have the highest sSFR. They are not extreme starbursts and can be found among normal main-sequence galaxies.
These conditions of high ionization can occur in galaxies with averaged  sSFR at cosmic dawn. Even some quiescent galaxies with lower-than-averaged sSFR can host these extreme ionization conditions.
This indicates that the line ratios mostly depend on the local conditions within star-forming regions, 
characterized by the ionization parameter. It depends on the local star-formation activity and the filling factor of dense gas,
regardless of the overall galaxy SFR or stellar mass.
At this point we may wonder whether there is any other global galaxy property that correlates with the local nebular conditions.

\subsection{Compact versus extended galaxies}
\label{sec:size}

In our sample, the diversity in line ratios is related to the different distribution of the SF regions within galaxies.
\Fig{maps} shows the  maps of [OIII] and \Halpha \ emission of two main-sequence galaxies with similar sSFR but different [OIII]/\Halpha \ ratios.
The left panels show a [OIII]-bright emitter with log([OIII]/\Halpha$)=0.15$.
The [OIII] emission is mostly concentrated in a compact central clump. 
The half-light area, $\pi {\rm R}_{50}^2$, corresponds to a half-light radius of only 0.5 kpc.
This compact morphology may be related to the fact that the galaxy is at the peak of a moderate SF burst. 
This may be driven by a compaction event \citep{DekelBurkert, Zolotov15}, 
in which a significant flow of gas within the galaxy reaches the galaxy center.
This leads to extreme conditions of star formation with a relatively high value of the averaged ionization parameter, ${\rm log \ U}=-2.7$. 
These nebular conditions boost the [OIII] luminosity with respect to \Halpha,
specially at the galaxy center, where the ionization parameter is the highest, ${\rm log \ U}=-2.5$ (bottom panel of  \Fig{maps}).

The right panels show a \Halpha-bright galaxy with log([OIII]/\Halpha$)=-0.17$.
In this case, the [OIII] luminosity is much more extended than in the previous example.
Its half-light radius is 1 kpc. 
However, the distribution is not smooth.
Most of the [OIII] emission is concentrated in a a few bright clumps extended over
the regions with the highest ionization parameter.
However, the nebular conditions are less extreme, with an averaged ionization parameter of ${\rm log \ U}=-3.3$. 
Under these conditions, \Halpha \ dominates over [OIII]. 

In conclusion, [OIII] emission is boosted in extremely compact SF clumps, where the ionization conditions can be very extreme with a very high ionization parameter. 
On the other hand, \Halpha \ dominates over [OIII] emission if the SF is distributed over several smaller clumps, where the ionization parameter is much lower.
As a consequence, \Halpha \ emission seems more extended than [OIII] in both examples (\Fig{maps}).
It covers the inter-clump medium where the ionization parameter is lower.
	  		 
\begin{figure}
	\includegraphics[width=\columnwidth]{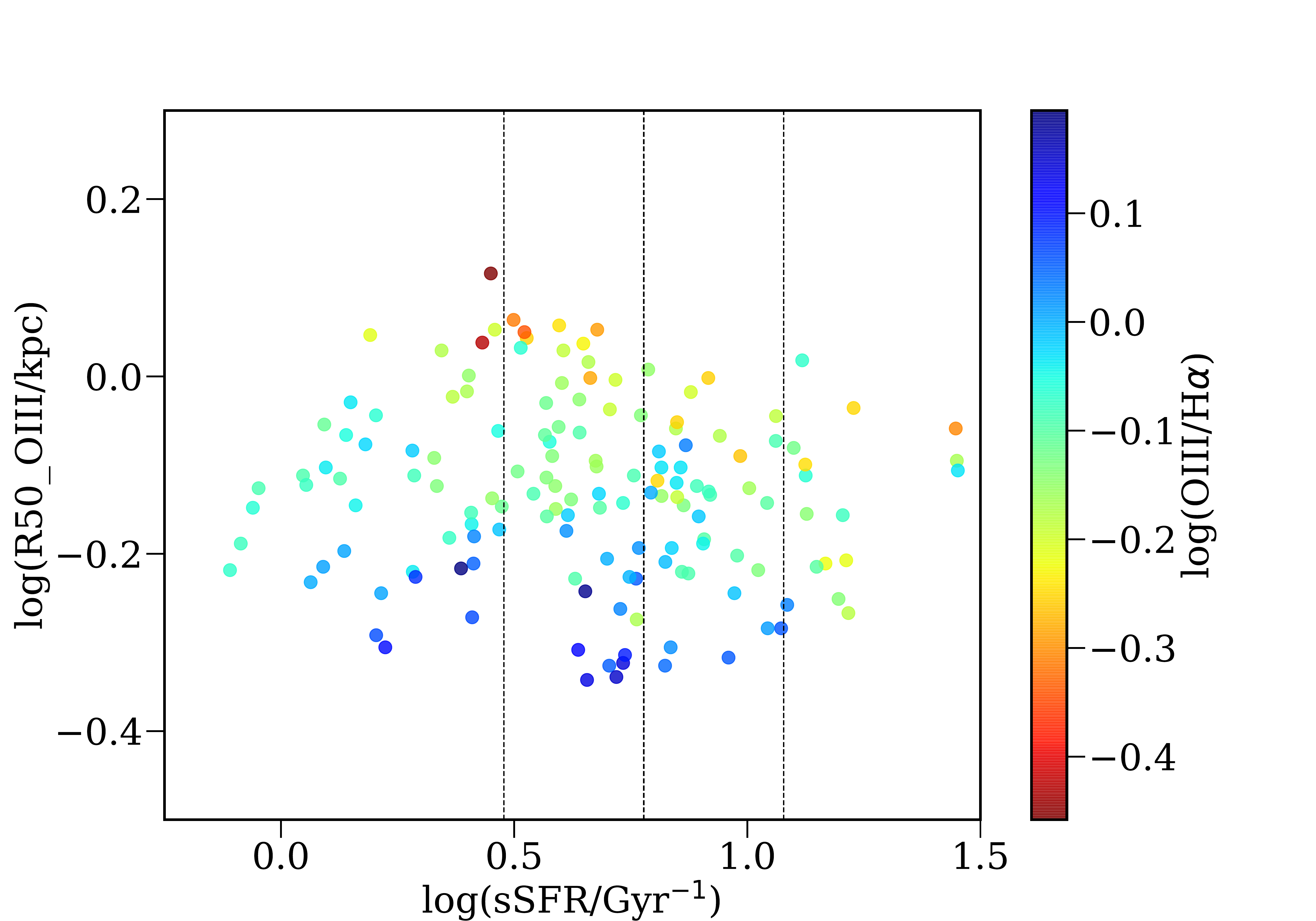}
	\includegraphics[width=\columnwidth]{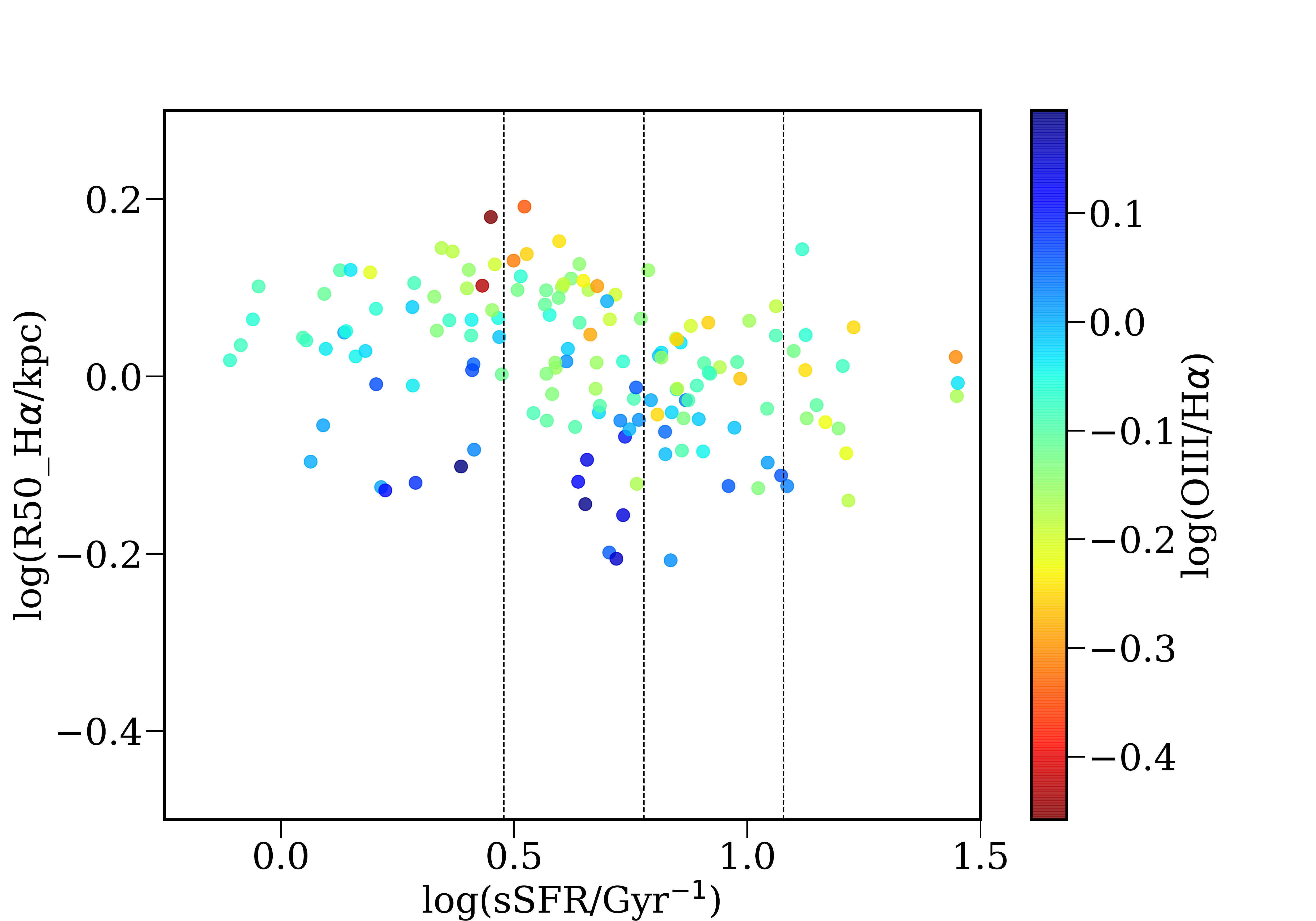}
		 \caption{Half-light radius versus sSFR for [OIII] (top) and \Halpha \ (bottom). 
		 The vertical lines mark the position of the star-forming main sequence and its scatter. 
		 The scatter at a fixed sSFR is mostly correlated with the line ratio. 
		 [OIII]-bright emitters are systematically more compact than \Halpha-bright galaxies.}	
	  \label{fig:R50}
\end{figure}

\Fig{R50} shows the half-light radius of the nebular emissions for the whole sample.
The diversity of sizes, even between galaxies of similar mass and SFR, is more evident.
It correlates with the [OIII]/\Halpha \ ratio and the ionization parameter. 
[OIII]-bright galaxies tend to be more compact than \Halpha-bright galaxies at any sSFR.
This is true even for galaxies below the main sequence.
Therefore, any observed sample of extreme [OIII] emitters will be systematically biased towards compact, smaller objects.
Many of these sources will be barely resolved by {\it JWST}, assuming a FWHM of around 0.12 arcseconds, which corresponds to a minimum half-light radius of  $\sim$300 pc at $z=6.5$.
The combination of [OIII]-bright and \Halpha-bright emitters will provide a more representative sample of galaxies at cosmic dawn.

The size of the \Halpha \ emission is significantly larger by 25\% (\Fig{R50}) . 
Therefore, it may be better resolved in future observations. FirstLight predicts many \Halpha \ emitters with effective sizes larger than 1 kpc at the end of reionization. NIRSpec IFU observations may be posible for galaxies with moderate sSFR at $z<6.5$.
	  
\section{Discussion and Conclusions}
\label{sec:summary}

We have used a sample of sub-L$_*$ galaxies with UV magnitudes, M$_{\rm UV}\simeq -19$ at $z\simeq6$, extracted from the FirstLight simulations \citep{PaperI} to study the diversity of galaxies at the end of the reionization epoch. The main results can be summarized as follows:
\begin{itemize}
\item
A factor $\sim$10 variation in the equivalent width of the [OIII]$\lambda$5007 line is mostly driven by a factor $\sim$40 variation in the specific SFR.
\item
Variations in nebular metallicity and ionization parameter within HII regions generate a dispersion in the equivalent width and [OIII]/H$\alpha$ \ line ratio at a fixed sSFR of $\sigma \simeq 0.1$ dex.
\item
OIII-bright (log([OIII]/\Halpha$)>0$) emitters have higher ionization parameters and/or higher metallicities than \Halpha-bright (log([OIII]/\Halpha$)<0$) galaxies.
 \item
According to the surface brightness maps in both [OIII] and \Halpha,  [OIII]-bright emitters are more compact than \Halpha-bright galaxies.
\item
\Halpha \ dominates over [OIII] if the star formation is distributed over extended regions. [OIII] dominates if the star formation is concentrated in large and compact clumps.
\item
The spatial extend of the \Halpha \ emission is significantly larger than that of the [OIII] emission.
\end{itemize}

These results indicates a large diversity in galaxy properties by the end of reionization.
Even galaxies with similar UV properties may have very different rest-frame visible emission lines.
Any compilation based only on [OIII] or \Halpha \ emitters may miss a significant and important population. 
[OIII]-bright emitters have on average compact emission regions with relatively high metallicity and/or high ionization parameters.
On the other hand, \Halpha-bright galaxies tend to be significantly more extended and their SF regions have less extreme nebular conditions. 

Different processes may transform  \Halpha-bright galaxies into [OIII]-bright emitters. 
For example, a large inflow of gas to the galaxy center, triggered by a compaction event \citep{DekelBurkert, Zolotov15} may generate a central starburst and change the conditions of the SF regions. Violent disk instabilities \citep{DSC, Ceverino10} produce giant clumps with similar gas conditions. The relevance of these processes at cosmic dawn remains to be explored in more detail in future works.

One of the caveats in the present analysis is the omission of radiative transfer effects from the intervening gas.
We plan to extend this analysis to more massive galaxies in bigger cosmological volumes and we will need to take into account the effect of dust attenuation. 
The calculation of the emission lines of this paper relies on particular models of nebular emission. They have their own limitations and assumptions about stellar binarity or rotation. However, the two models used in this paper give very similar line luminosities.  This assures us that the diversity reported in this paper is independent of a particular model and it is a strong prediction of the FirstLight simulations. Future observations with {\it JWST} and next generation telescopes will be able to test these results.

	  
\section*{Acknowledgements}

We thank the anonymous referee for comments that improve the quality of this paper.
We acknowledge stimulating discussions with Xiangcheng Ma, Sune Toft, and Luis Colina.
DC is a Ramon-Cajal Researcher and is supported by the Ministerio de Ciencia, Innovaci\'{o}n y Universidades (MICIU/FEDER) under research grant PGC2018-094975-C21.
This work has been funded by  the ERC Advanced Grant, STARLIGHT: Formation of the First Stars (project number 339177). 
RSK and SCOG also acknowledge support from the DFG via SFB 881 `The Milky Way System' (sub-projects B1, B2 and B8) and SPP 1573 `Physics of the Interstellar Medium' (grant number GL 668/2-1) and KL 1358/19-2.
AF acknowledges the support from grant PRIN MIUR2017-20173ML3WW\_001.
The authors gratefully acknowledge the Gauss Center for Supercomputing for funding this project by providing computing time on the GCS Supercomputer SuperMUC at Leibniz Supercomputing Centre (Project ID: pr92za).
The authors acknowledge support by the state of Baden-W\"{u}rttemberg through bwHPC.
We thank the BPASS team for sharing their database of SSPs and emission lines. This work made use of the v2.1 of the Binary Population and Spectral Synthesis (BPASS) models as last described in \cite{Eldridge17}. 


\section*{Data Availability}

The data underlying this article are available in the FirstLight database, at \url{http://odin.ft.uam.es/FirstLight} or
will be shared on reasonable request to the corresponding author.



\bibliographystyle{mnras}
\bibliography{FLIV_v5} 



\appendix

\section{The [OIII]/H$\beta$ line ratio}

\begin{figure}
	\includegraphics[width=\columnwidth]{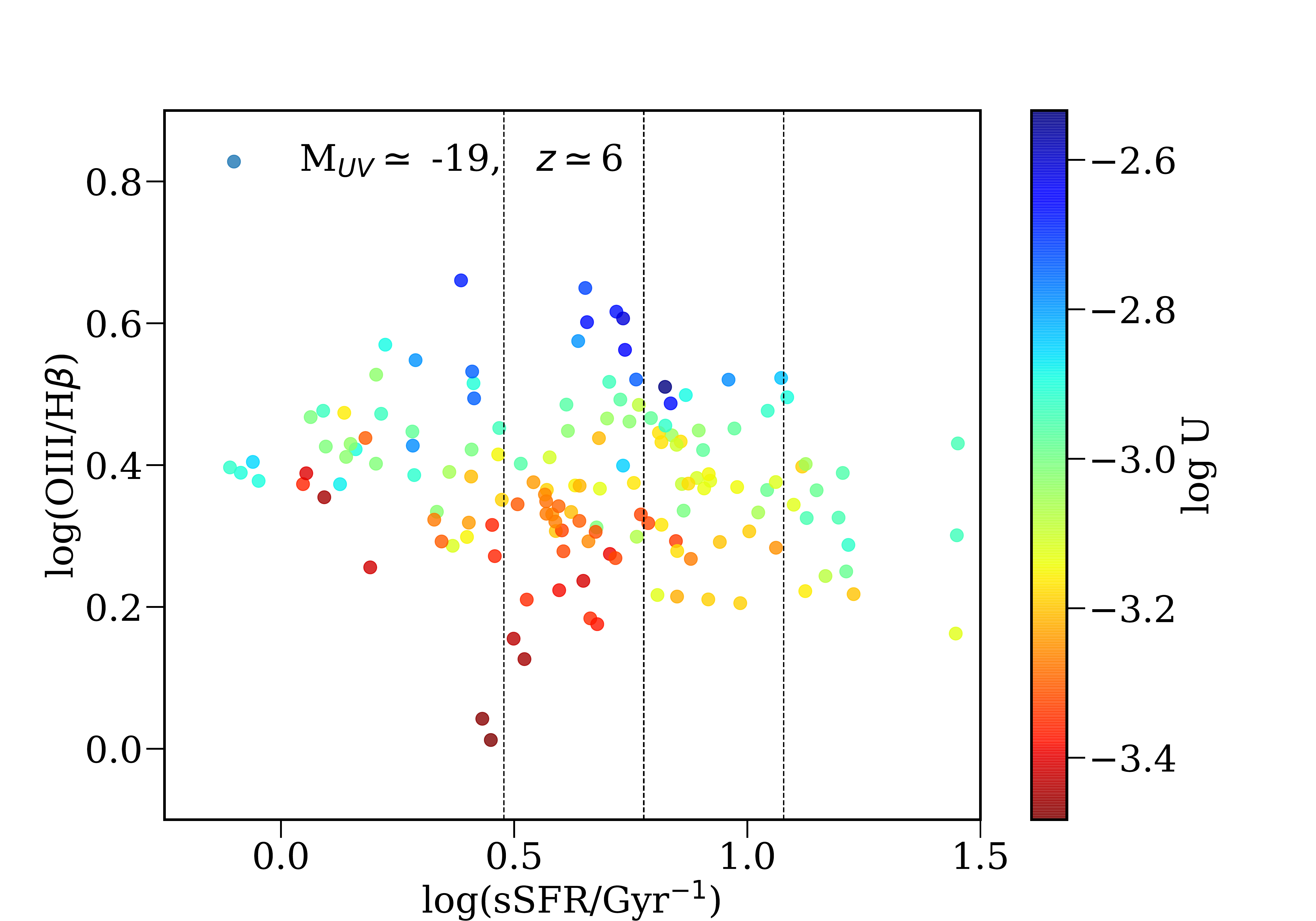}	
		 \caption{[OIII]/H$\beta$ versus sSFR. There is a systematic offset in comparison with the [OIII]/H$\alpha$ ratio due to the intrinsic ratio of 2.85 between \Halpha \ and H$\beta$.}
	  \label{fig:OHbeta}
\end{figure}

This paper was mostly focused on the comparison between [OIII]$\lambda$5007 and \Halpha, the two brightest lines in the rest-frame visible.  However the [OIII]/H$\beta$ line ratio is also commonly used at low redshift. \Fig{OHbeta} provides an analog to \Fig{OHalpha} and it shows that the [OIII]/H$\beta$  ratio is systematically lower by a factor of 2.85, the intrinsic ratio between \Halpha \ and H$\beta$. This overall offset is expected because dust attenuation is very low in these sub-L$_*$ galaxies.

\bsp	
\label{lastpage}
\end{document}